# STUDIES

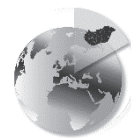

# The relationship between social innovation and digital economy and society

**Szabolcs Nagy**
Faculty of Economics,
University of Miskolc,
Hungary
E-mail:
nagy.szabolcs@uni-miskolc.hu

**Mariann Veresné Somosi**
Faculty of Economics,
University of Miskolc,
Hungary
E-mail: szvvsm@uni-miskolc.hu

The information age is also an era of escalating social problems. The digital transformation of society and the economy is already underway in all countries, although the progress in this transformation can vary widely. There are more social innovation projects addressing global and local social problems in some countries than in others. This suggests that different levels of digital transformation might influence the social innovation potential. Using the International Digital Economy and Society Index (I-DESI) and the Social Innovation Index (SII), this study investigates how digital transformation of the economy and society affects the capacity for social innovation. A dataset of 29 countries was analysed using both simple and multiple linear regressions and Pearson's correlation. Based on the research findings, it can be concluded that the digital transformation of the economy and society has a significant positive impact on the capacity for social innovation. It was also found that the integration of digital technology plays a critical role in digital transformation. Therefore, a country's progress in digital transformation is beneficial to its social innovation capacity. In line with the research findings, this study outlines the implications and possible directions for policy.

## Introduction

We live in an age of environmental and social problems. According to the ‚Global Shapers Annual Survey 2017', one of the latest reports of the World Economic Forum that reflects the view of millennials, the 12 most concerning world issues are climate change, destruction of the natural environment, extensive conflicts and wars, inequality (income and discrimination), poverty, religious conflicts, accountability and transparency of governments, corruption, food and water safety, lack of education, security and prosperity, and the lack of economic opportunities and unemployment (WEF 2017). The main concerns revealed by this study are in line with earlier research results for Hungary, as Hungarians addressed unemployment, ethnic conflicts, poverty, armed conflicts/wars, difficult access to quality education, environmental pollution, and health issues as the most pressing social problems (Nagy 2012). Social innovations attempt to find solutions to these issues.

Some researchers (Howaldt–Schwarz 2010, Van der Have–Rubalcaba 2016, Kocziszky et al. 2017) consider social innovation as a relatively new concept, expected to spread quickly because of a current paradigm shift in innovation research. However, social innovation has always been present in the development of societies in an invisible, implicit way, because development itself cannot be imagined in any society without innovation and new types of responses to existing problems. Nevertheless, social innovation is a buzzword in the debate on the changes required to meet challenges the EU currently faces (Misuraca–Pasi 2019).

According to DSI (2021a), there are 2,314 organizations and 1,521 projects working on digital social innovation across Europe. However, the distribution of organisations and projects is uneven (DSI 2021b). The United Kingdom, Spain, Italy, France, and the Netherlands are the leading countries in terms of the number of social innovation projects. At the other extreme, Bosnia and Herzegovina, Cyprus, and Lithuania are where we hardly ever find such projects. Consequently, it can be assumed that the difference in the number of digital social innovation (DSI) projects and organisations can be explained by the different levels of digitalisation across countries. Additionally, these countries' digitalisation advancement is quite diverse, as shown in their ranking in the Digital Economy and Society Index (European Commission 2018, 2020).

Previous studies failed to address how digital transformation influences social innovation; therefore, it is timely and critical to investigate the relationship between digital performance and competitiveness and the social innovation capacity of a country. It is hypothesised that higher digital performance and competitiveness positively influence social innovation capacity (H1). The main objective of this study is to determine how the digital transformation of society and the economy affect the capacity for social innovations. To achieve this objective, the relationship between the two indices, the Social Innovation Index (SII) and the International Digital Economy and Society Index (I-DESI), will be examined. Furthermore, to add to the overall body of academic literature in the field of social innovation, this study aims to analyse the relationship between each dimension of the I-DESI and the SII.

## Background

### Social innovation

No commonly agreed definition of social innovation can be found in existing literature. According to the Centre for Social Innovation at Stanford Graduate School of Business, social innovation is a process that supports social progress by developing and implementing effective solutions to challenge social and environmental issues (Centre for Social Innovation 2021). Social innovations are new responses to pressing social demands, and new ideas related to products, services, and models that address social needs more effectively than alternatives and create new social relationships or collaborations (BEPA 2010).

Social innovations are related to all social and societal demands and challenges, where any innovation (renewal) can lead to improvement through novelty and development (Mulgan 2006). They are innovative activities and services that meet social needs. Social innovation is a novel solution to a social problem (Phills et al. 2008) and is mostly related to organisations with clear social purposes (Mulgan et al. 2007). The main objective of social innovation is to improve human well-being (BEPA 2010). It serves the interest of the whole society rather than the individuals' and boosts society's capacity to act. Social innovations demand and bring about changes in a particular social system (Westley–Antandze 2010).

Social innovation is as important as traditional innovation. The latter often causes social problems that can only be solved through social innovation. Therefore, traditional and social innovation should be used together (Kocziszky et al. 2017). There is a growing need for social innovations due to the negative societal and economic consequences of the COVID-19 pandemic (Bacq–Lumpkin 2020).

Although there is a strong demand from the policy-making institutions to measure social innovation, only a few measuring constructs are available due to the lack of consensus on the interpretation of social innovation, its determining factors, and the most appropriate methodologies. According to Unceta et al. (2016), three different approaches can be used: the individualist, the organisational, and the regional/national. The Global Entrepreneurship Monitor is related to the individualist approach, whereas the pilot Project RESINDEX (SINNERGIAK Social Innovation 2013) is associated with the organisational approach. The Regional Social Innovation Index (RESINDEX) developed by Unceta et al. (2016) is also based on this approach. Third, the EU-wide TEPSIE project is focused on developing social innovation indicators at the macro level, using the regional/national approach (Krlev et al. 2014). The SII developed by the EIU (2016) also fits this line.

## Digitalisation and digital transformation

Digital transformation of the economy and society is underway in every country in the world. According to Katz (2017), digitisation can be considered a transformation triggered by the massive adoption of digital technologies that generate, process, share, and transfer information. Four waves of digital transformation, driven by technological progress and innovation diffusion, can be distinguished. The emergence of management information systems to analyse business performance and telecommunication technologies was the first wave. This was followed by the second wave, associated with the diffusion of the internet and its corresponding platforms. The third wave is adopting a range of advanced technologies, including the internet of things (IoT), robotics, sensors, artificial intelligence, big data, and analytics to support decision-making and problem-solving. The latest wave of digitisation is closely related to productivity improvements but may also have significant benefits on social welfare, especially on environmental issues and sustainable development, which are the main fields of social innovation (Katz 2017).

Schwab (2016) also identified four revolutions, profound changes in the economic and social systems caused by new technologies. The current, fourth industrial revolution is related to and can be characterised by the widespread use of artificial intelligence, machine learning, and the IoT. This is also an era of digital transformation, where digital technologies are used to change all aspects of the business world. COVID-19 gave another strong boost to digitalisation in many fields and drastically transformed how we work and live. This might be the beginning of a new fifth era, which can be called the age of 'forced digital innovation' due to its unintentional characteristic. Digitalisation has accelerated as a result of the COVID-19 pandemic and has changed the way we work (Lipták 2021), and contributes to sustainability (Szalmáné Csete–Buzási 2020) and creating smart cities (Mitrofanova et al. 2021).

The latest report by the OECD (2020) summarises the constructs developed by intergovernmental and international organisations that comprise indicators to measure digital transformation at the macro level. The first is the Digital Economy and Society Index (DESI 2020) by the European Commission. The second one is the 'Measuring digital development: Facts and figures series' by the International Telecommunication Union (ITU 2020) which presents the state of digital development across its 196 member countries. The third is the OECD Going Digital Toolkit, with 33 core and several complementary indicators (OECD 2021). The next one is the 'Digital Economy' portal of the United Nations Conference on Trade and Development (UNCTAD 2020) on information and communication technologies (ICT) trade and ICT use in businesses. The fifth construct, UNESCO's ROAM-X framework, assesses internet universality in four categories (rights, openness, accessibility, multi-stakeholder) with a set of 303 indicators. It provides a holistic diagnosis of internet policies, the digital environment, and the structural causes of digital inequalities (UNESCO 2021). In partnership with the Institute of the Information Society, the World Bank Group developed the Digital Economy Country Assessment (DECA) to evaluate the preparedness for digital adaptation at the country and regional levels (World Bank Group 2018). The digital transformation of society and the economy is so accelerated nowadays that existing metrics and measurement tools need to be revised and replaced (OECD 2019). The I-DESI can be considered a step forward to meet this requirement (European Commission 2021).

### Digital social innovation

According to European Commission–Bria (2016), DSI is a new, digital approach to addressing social issues. DSI is both a social and collaborative innovation. The main emphasis is on collaboration among innovators, users, and communities to use digital technologies to solve social problems. The co-creation of knowledge and solutions for a wide range of social needs is also important in this process. Similarly, Milwood–Roehl (2019) consider DSI a type of social innovation that uses new technologies to address a broad range of social problems.

Digital social innovation has recently become more significant due to its critical role in dealing with the adverse social consequences of the challenges caused by COVID-19 (Bapuji et al. 2020). Social innovation opportunities for digital brokering and digitised services have been addressed through improvised venturing, rapid pivoting, and pro-social product extension (Scheidgen et al. 2021).

Huh–Kim (2019) analysed the influencing factors of multidisciplinary DSI in OECD member countries and found high levels of democracy and e-participation, high gross domestic product (GDP) and business-friendly environments, high social expenditure, and high level of ICT development and patent applications have a positive effect on digital social innovation.

### The link between digitalisation, digital transformation, and social innovation

The growing number of ICT-based innovative solutions positively impacts social innovation potential, as ICT often plays a key role in the development of platforms supporting innovative partnerships (Misuraca–Pasi 2019). The way ICT is used by individuals, businesses, and governments significantly influences the ability of countries to benefit from the social, economic, and environmental impact of ICT ecosystems (Vrontis et al. 2021). An ecosystem must be developed that provides community members with the possibility, means, opportunities, and authority to pursue their social utility objectives, to support social innovation initiatives and social enterprises (Chi et al. 2020).

Digitalisation and social entrepreneurship contribute to national well-being (Torres–Augusto 2020). Digitalisation positively impacts the society. This makes it easier for citizens to access public services and contribute to higher employment and economic growth (Galindo-Martín et al. 2019). The authors also found that digitalisation positively affects national well-being only if some criteria are met: a good educational system, a philanthropic financial system, and adequate governance must also be present. In contrast, social entrepreneurship does not affect the achievement of national well-being.

Shenkoya–Dae-Woo (2019) examined the impact of IoT on Japanese society and found that the diffusion of the IoT has a positive relationship with social innovation in Japan, and IoT positively affects Japanese society, although it causes only an incremental change.

## Operationalisation

To meet the research objectives, the relationship between the I-DESI, and the SII was analysed. The SII measures the capacity for social innovation in 45 countries. I-DESI quantifies and compares the digital economy performance of 45 countries: 27 EU member states with 18 other countries (Australia, Brazil, Canada, Chile, China, Iceland, Israel, Japan, Mexico, New Zealand, Norway, Russia, Serbia, South Korea, Switzerland, Turkey, the United Kingdom, and the United States).

### Social Innovation Index (SII) and ranking

The SII was devised and constructed by the Economist Intelligence Unit (EIU) research team, sponsored by the Nippon Foundation in 2016 (EIU 2016). The SII aims to investigate how countries can encourage and enable social innovation. This construct measures the capacity for social innovation in 45 countries, including G20 and OECD nations and some developing countries from different geographical regions. It includes seven quantitative data points and ten qualitative scores from EIU analysts, grouped into four pillars. The four pillars of SII represent the capacity of a country to develop social innovation. Data points within each pillar are normalised from a scale between 0 and 100, where 0 represents the worst and 100 the best, and then assigned weights. Scores for each pillar are then calculated and normalised out of 100. Each pillar is given a different weight in the overall score, out of 100 (EIU 2016).

The four pillars, their weights, and the indicators that comprise the components are as follows:
1. Policy and institutional framework (weight: 44.44%)
   – Existence of national policy on social innovation (25%)
   – Social innovation research and impact (20%)
   – Legal framework for social enterprises (20%)
   – Effectiveness of system in policy implementation (20%)
   – The rule of law (15%)
2. Financing (weight: 22.22%)
   – Availability of government financing to promote social innovation (50%)
   – Ease of getting credit (25%)
   – Total public social expenditure (25%)
3. Entrepreneurship (weight: 15%)
   – Risk-taking mind-set (25%)
   – Citizens' attitude towards entrepreneurship (25%)
   – Ease of starting a business (25%)
   – Development of clusters (25%)

   4.  Society (weight: 18.33%)
       – Culture of volunteerism (20%)
       – Political participation (20%)
       – Civil society engagement (20%)
       – Trust in society (20%)
       – Press freedom (20%)

*The existence of national policy on social innovation,* which is the most important indicator of the first pillar, refers to the existence of a government-led national policy to encourage social innovation. Maximum points (2) are given where there exists a government strategy on promoting social innovation entrepreneurship, and zero (0) where no such strategy exists.

*Social innovation research and impact* measures the existence of government-led data collection and policy needs to support social innovation. Three (3) points are given where the government regularly collects information on social enterprises and entrepreneurs with data made public and zero (0) points in countries with no such research. *The legal framework for social enterprises* is the existence of specific regulatory frameworks for social enterprises, social entrepreneurs, and other social innovation businesses. Two (2) points where legal frameworks exist and are widely used and zero (0) where no such frameworks exist. *Effectiveness of system in policy implementation* measures the effectiveness of policy implementation and execution on a five-point scale (5=very high, 1=very low). *Rule of law*, which is the least important indicator composing the Policy and institutional framework pillar, measures the transparency and fairness of the legal system on a five-point scale (5=very high/fair, 1=very low/unfair).

The first and most important indicator of the second pillar, financing, is the *availability of government financing to promote social innovation* that measures the availability and ease of use of financing mechanisms such as social innovation funds, government grants, social impact bonds, and business incubators. A maximum of seven (7) points can be achieved, where all mechanisms are available and easy to access and zero (0) where there is no such availability. *Ease of getting credit* measures the degree to which collateral and bankruptcy laws protect the rights of borrowers and lenders and thus facilitate lending (12=very high, 0=non-existent). *Total public social expenditure* in terms of percentage of GDP measures the government's social expenditure in the form of cash benefits, direct in-kind provision of goods and services, and tax breaks with social purposes.

The third pillar of the SII comprises four equally important indicators. The first is the *risk-taking mind-set*. It measures the percentage of the population aged 18–64 years with positive perceived opportunities, indicating that fear of failure would prevent them from setting up a business. The second indicator, *citizens' attitude towards entrepreneurship*, measures the percentage of the population aged 18-64 years who agree with the statement that in their country, most people consider starting a business as a desirable career choice. *Ease of starting a business* scores the levels of regulation involved in setting up new private businesses on a five-point scale (5=very high, 1=very low). *Development of clusters* measures the extent to which there are well-developed and deep clusters (geographic concentrations of firms, suppliers, producers of related products and services, and specialised institutions) in a particular field on a seven-point scale (7=widespread in many fields, 1=non-existent).

The fourth pillar, Civil Society, comprises five equally important indicators, each having 20% weight. *Culture of volunteerism* shows the average percentage of people in each country who donate money, volunteer or help a stranger. *Political participation* measures citizens' willingness to participate in public debates, elect representatives, and join political parties. Maximum points (10) indicate the highest level of participation, whereas zero (0) indicates the lowest participation. *Civil society engagement* in terms of percentage of the population is the proportion of respondents who are members (active or inactive) of a humanitarian or charitable organisation. *Trust in society* shows the proportion of respondents who answered 'most people can be trusted' in the World Values Survey, European Social Survey, Latinobarómetro, and Global Barometer Study. *Press freedom* refers to the level of freedom available to journalists based on the results of the World Press Freedom Index on a hundred-point scale (100=best, 0=worst).

Although the SII has some limitations, i.e., not all the countries in the world are included, interpretation of the results can only be possible in the domestic context, a systematic measurement of private-sector inputs is missing, and it does not measure the outputs of social innovation, this is a useful tool for measuring social innovation capacity.

In the SII 2016 (Figure 1), the United States had the highest social innovation potential, followed by the United Kingdom (77.3) and Canada (75.7). Overall, the U.S.A. scored 79.4 out of 100, and its leading position resulted from its high scores achieved in each of the four pillars. Denmark, Belgium, New Zealand, France, Germany, Sweden, and Switzerland can be found in the top ten highly developed countries. Each of these countries has a very high per-capita income, human development indicators, and stable democratic governance. The Philippines (27.6), Paraguay (28.1), and Saudi Arabia (30.2) were the three worst performing countries in this ranking. The low scores achieved by these countries reflect their weak social innovation capacity. Hungary is not ranked in the SII 2016; therefore, in this study, an attempt will be made to predict its capacity for social innovation using Hungary's overall I-DESI score.

Figure 1

Social Innovation Index 2016 rankings

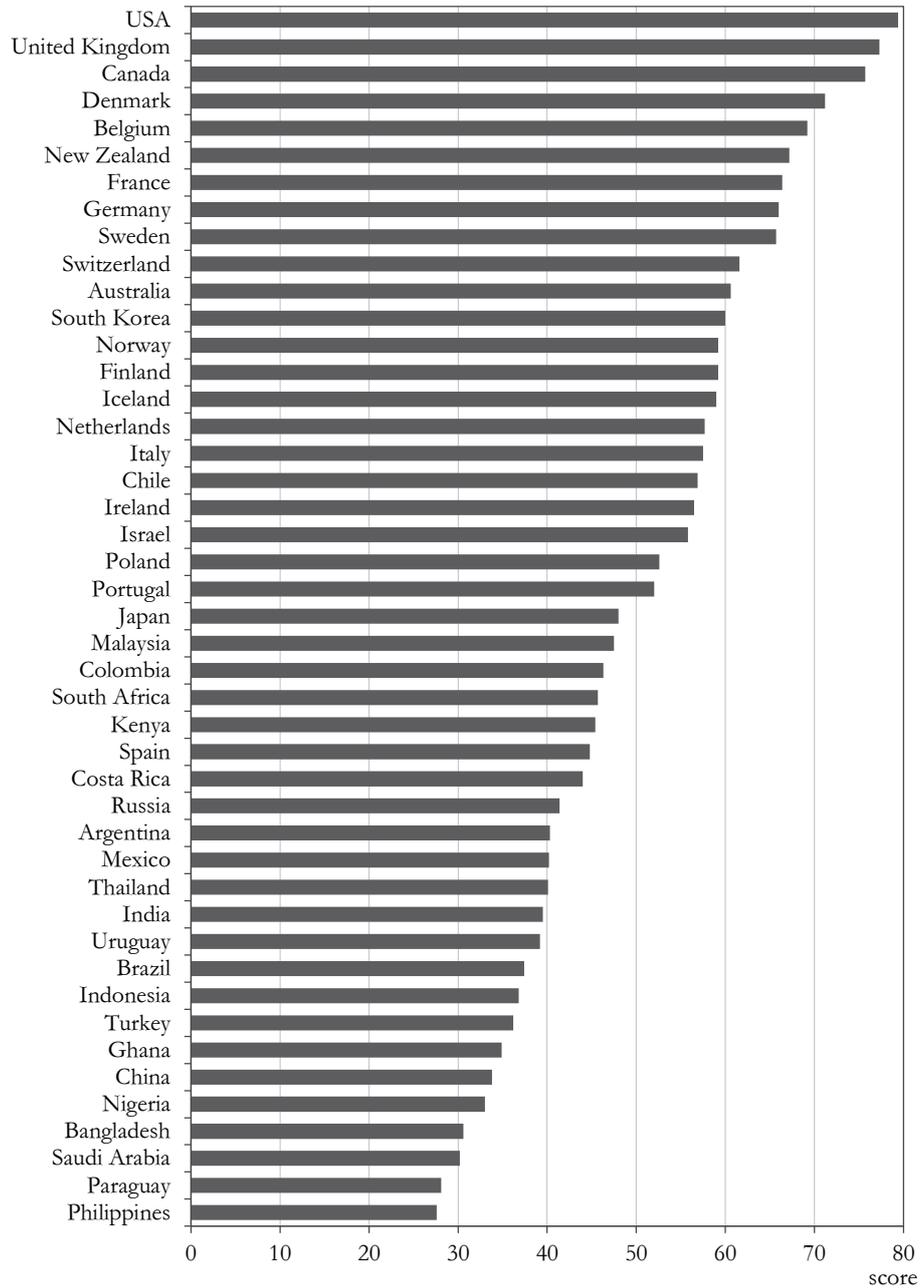

*Source:* EIU (2016).

### International Digital Economy and Society Index (I-DESI) and ranking

The I-DESI is a composite index, an extended and advanced version of the EU Digital Economy and Society Index (DESI). It is suitable for trend analysis and to compare the digital performance of 45 countries, including 27 EU member states and 18 non-EU countries. As far as the latter countries are concerned, six are from Europe, five from Asia, five from the Americas, and two from Australasia (European Commission 2021). I-DESI eliminates one of the major drawbacks of DESI, namely, the difficulties in comparing time-series performances due to the frequent changes in the measurement factors (Bánhidi et al. 2020).

I-DESI is composed of 24 relevant indicators of the digital performance and competitiveness of the countries across five principal dimensions: connectivity, human capital, use of internet (services by citizens), integration of digital technology (by businesses), and digital public services (European Commission 2021).

The overall I-DESI is calculated as the weighted average of the five main dimensions. Connectivity and human capital are the two equally important dimensions with a weighted score of 25% each, followed by the integration of digital technology (20%). The two least important dimensions in this index are the use of the internet (15%) and digital public services (15%). Therefore, the I-DESI score can be calculated for a country using the following formula:

$$I - DESI = Connectivity * 0.25 + Human\ Capital * 0.25 + Use\ of\ Internet * 0.15 \\ + Integration\ of\ Digital\ Technology * 0.2 + Digital\ Public\ Services * 0{,}15$$

The connectivity dimension quantifies the deployment of broadband infrastructure and its quality. Fast and ultrafast broadband networks and access to services using these networks are key factors in competitiveness. It includes six indicators: fixed broadband coverage, fixed broadband take-up, 4G coverage, mobile broadband take-up, fixed broadband speed, and fixed broadband price. The human capital dimension, made up of five indicators, measures the digital skills of users. Basic skills, at least basic software (individuals who have written computer code in the last 12 months), Telecomm FTEs (full-time equivalent telecommunication employees), and ICT graduates contribute to this dimension. Various online activities, including social networking, making video calls, shopping, and banking online, contribute to the use of internet services dimension, which contains six indicators: internet users, fixed broadband traffic, video calls, social networks, banking, and shopping.

Figure 2

International Digital Economy and Society Index ranking

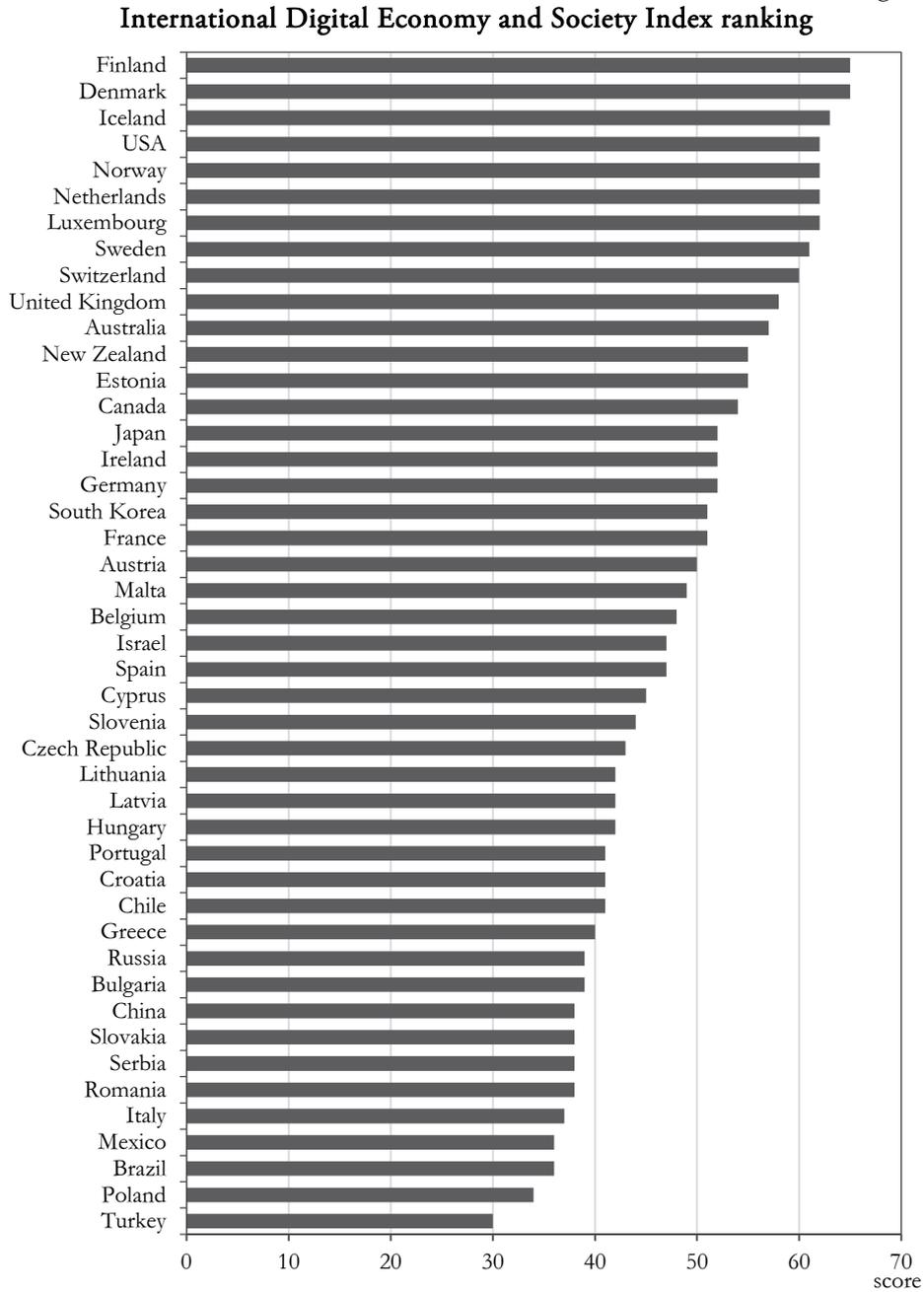

*Source:* European Commission (2021).

The integration of digital technology (IDT) by businesses refers to the digitisation of businesses and e-commerce. Adaptation of digital technologies positively impacts efficiency and cost reduction; furthermore, digitisation of services makes it possible for digitally savvy companies to improve customer experience. Four indicators contribute to IDT: availability of latest technologies, firm-level technology absorption, small and medium-sized enterprises (SMEs) selling online, and secure internet servers. Finally, the digital public services dimension focuses on the digitisation of public services. Modern and advanced societies are characterised by providing a high level of digital public services to citizens and businesses. This dimension comprises only three indicators: participation index, online service completion, and open data (European Commission 2021).

Figure 2 provides an overview of the performance scores across all dimensions of the I-DESI. The leading country in this study was Finland, with an average score of 65.0 across the four years (2015–2018) of this study. Finland was also the leading country in the 2019 and 2020 EU DESI studies. The second was Denmark (64.8), and the leading non-EU country, which came third of all 45 countries studied, was Iceland (62.7).

At the other end of the ranking, Brazil, Poland and Turkey can be found. These are the least developed economies and societies in the world digitally. Hungary ranks 30th out of 45 countries in the middle of this ranking, implying a medium level of digital development.

## Research methodology and sample

To analyse the relationship between digital economy performance and the capacity for social innovation, only countries that have been involved in both the I-DESI 2016 and SII 2016 rankings were investigated. Consequently, 29 countries were included in the analysis. The raw data for analysis are presented in Appendix (Table A1). The sample comprised 13 European countries and 16 non-EU countries. In Table A1, the countries are listed according to their social innovation capacity. The relatively small sample size of 29, which is still significantly higher than it would be if the analysis were based on the DESI scores of 13 countries, is one of the limitations of this study.

A simple linear regression was used to test the working hypothesis, which assumes that higher digital performance and competitiveness positively influence social innovation capacity (H1). The SII score was the dependent variable in simple linear regression, and the I-DESI score was the independent variable. Multiple linear regression was applied to investigate the effects of the five I-DESI components on the capacity for social innovation. Principal component analysis (PCA) was used to identify underlying patterns. The relationships between the dimensions of the SII and I-DESI were investigated using Pearson's correlations.

## Results and discussion

To test the conditions for linear regression data cleaning was applied first. Descriptive statistics of the variables were examined to detect erroneous values (Table 1). Missing and/or erroneous values were not included in the raw dataset.

Table 1

### Descriptive statistics

| | SII | Policy and institutional framework | Financing | Entre-preneur-ship | Society | I-DESI | Con-nec-tivity | Human capital | Use of the internet | Integration of digital technology | Digital public services |
|---|---|---|---|---|---|---|---|---|---|---|---|
| Valid | 29 | 29 | 29 | 29 | 29 | 29 | 29 | 29 | 29 | 29 | 29 |
| Missing | 0 | 0 | 0 | 0 | 0 | 0 | 0 | 0 | 0 | 0 | 0 |
| Mean | 57.534 | 56.062 | 58.417 | 59.303 | 58.576 | 49.103 | 59.759 | 39.690 | 45.517 | 44.379 | 56.310 |
| Std. deviation | 12.202 | 16.210 | 14.540 | 8.118 | 18.714 | 10.520 | 9.425 | 11.604 | 14.339 | 12.448 | 16.123 |
| Minimum | 33.800 | 28.800 | 36.900 | 44.800 | 24.000 | 26.000 | 40.000 | 19.000 | 19.000 | 19.000 | 24.000 |
| Maximum | 79.400 | 86.600 | 82.000 | 76.200 | 88.300 | 65.000 | 72.000 | 62.000 | 68.000 | 62.000 | 80.000 |

After checking the boxplots, it was concluded that there were no outliers in the database. The simple linear regression condition of independence was fulfilled because one country does not affect the digital development of the other country, that is, observations are independent of each other. The assumption for the scales was also met due to the use of normalised scales ranging from 0 to 100. To test normality, the Shapiro-Wilk test was applied. Both variables are normally distributed, as the P-values did not show a significant relationship (Shapiro-Wilk [SII]=0.965, P [SII]=0.427; Shapiro-Wilk [I-DESI]=0.945, P [I-DESI]=0.135). The scatter plot residuals vs. the predicted and standardised residual histogram (Figure 3) imply that the linear relationship and homoscedasticity conditions are met.

Figure 3
**Scatter plot residuals vs. the predicted and standardised residual histogram**

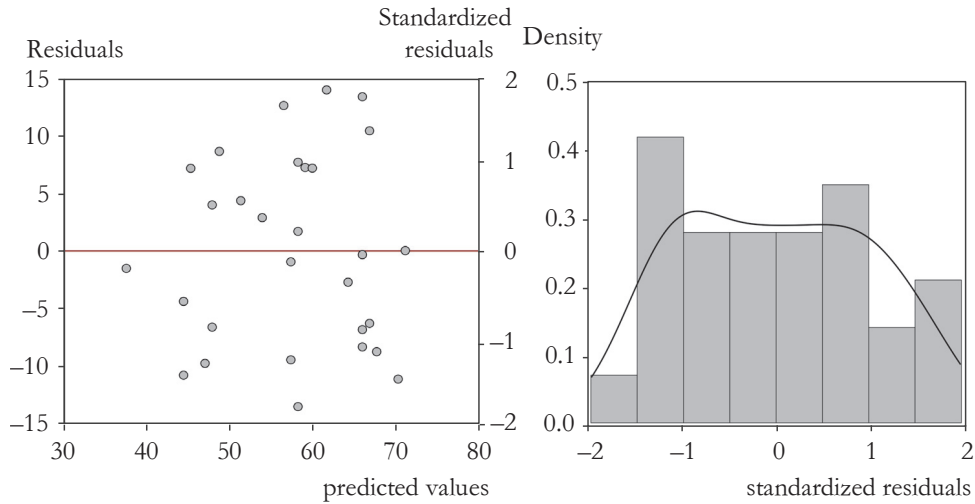

Durbin-Watson's d-value of 2.351, close to the ideal value of 2.0, indicates no disturbing autocorrelation detected in the sample. Figure 4 shows the countries' positions based on their innovation capacity (SII) and the development of the digital economy and society (I-DESI). This suggests that more digitally advanced countries have a greater capacity for social innovation.

Figure 4
**Scree-plot of countries based on their scores in SII and I-DESI**

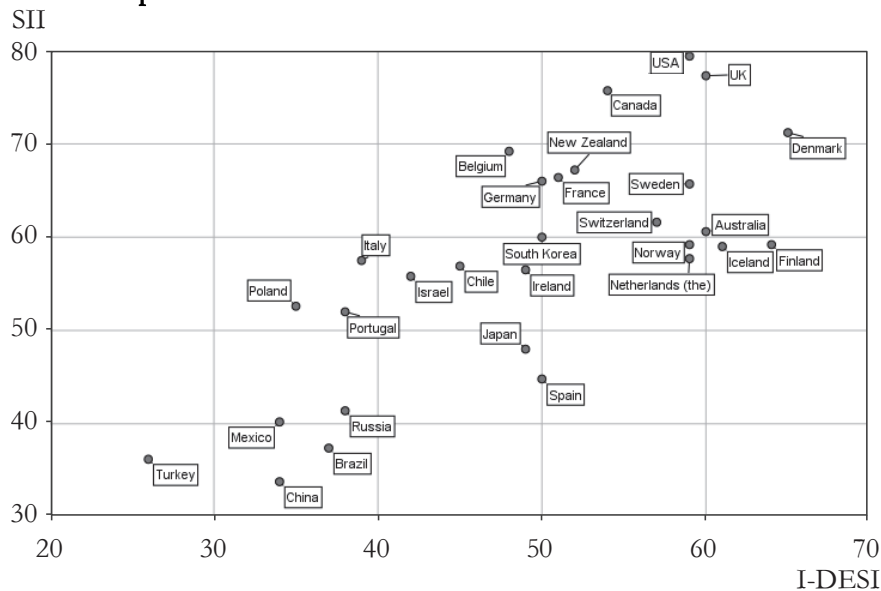

The ANOVA results show that the model is significant, i.e. generalisable to the population (regression sum of squares=2280.665, df=1, mean square=2280.665, F=32.611, p< .001).

As shown in Table 2, the $R^2$ value of 0.547 indicates that the resulting model is robust, with 54.7% of the variance in the output variable (SII) explained by the predictor variable (I-DESI). Pearson's correlation coefficient between the two variables (R=0.74) also suggests a strong relationship between social innovation capacity and progress toward the digital economy and society. The adjusted $R^2$ indicates that the model's explanatory power for the population is still very high (53.0%).

Table 2

### Linear regression model summary (SII and I-DESI)

| Model | R | $R^2$ | Adjusted $R^2$ | RMSE | Durbin-Watson | | |
|---|---|---|---|---|---|---|---|
| | | | | | Auto-correlation | Statistic | p |
| $H_0$ | 0.000 | 0.000 | 0.000 | 12.202 | –0.165 | 2.214 | 0.559 |
| $H_1$ | 0.740 | 0.547 | 0.530 | 8.363 | –0.233 | 2.351 | 0.338 |

The social innovation capacity of a country as a function of its digital development can be defined by the model as follows (Table 3):

$$SII = 15.408 + 0.858 \times I\text{-}DESI$$

The standardised beta coefficient suggests that I-DESI significantly and positively influences social innovation capacity. Therefore, H1 is accepted.

Table 3

### Coefficients (SII and I-DESI)

| Model | | Unstandardised | Standard error | Standardised | t | p |
|---|---|---|---|---|---|---|
| $H_0$ | (Intercept) | 57.534 | 2.266 | | 25.392 | <0.001 |
| $H_1$ | (Intercept) | 15.408 | 7.539 | | 2.044 | 0.051 |
| | I-DESI | 0.858 | 0.150 | 0.740 | 5.711 | <0.001 |

Based on this result, which is in line with those of Misuraca–Pasi (2019) and Vrontis et al. (2021), it can be concluded that *digital performance and competitiveness of the economy and society has a significant effect on the capacity for social innovation. A country's progress in digitalisation is beneficial to its social innovation capacity.*

Using the above formula, Hungary's capacity for social innovation can be predicted as follows:

$$SII (Hungary) = 15.048 + 0.858 \times 42 = 51.084$$

The predicted value is closest to the social innovation score of Portugal (52.0), suggesting that Hungary is very similar to Portugal regarding its potential for social innovation. This result is in line with those of Kolosi–Szivós (2018), which suggest that Hungary is closest to Portugal among the EU countries in many respects.

Multiple linear regression was used to investigate the impact of I-DESI components on social innovation capability. The dependent variable in the theoretical model was SII, whereas the independent variables were connectivity, human capital, use of the internet, integration of digital technology, and digital public services. As a rule of thumb, ten observations per variable and a minimum sample size of 30 is desirable in multiple linear regression (Hair et al. 2014), which suggests that the sample size of 29 is nearly acceptable and indicates another limitation of this research. This also implies that a maximum of two prediction variables should be allowed in the final model. Boxplots confirmed that no outliers were found in the variables. To detect multiple outliers: Cook's distances were calculated. The empty case-wise diagnostics table confirmed that there are no multiple outliers in the dataset. However, Pearson's correlation detected collinearity between the integration of digital technology and the use of the internet (Table 4).

Table 4

### Pearson's correlations

| Variable | | SII | Connec-tivity | Human capital | Use of the internet | Integ-ration of digital techno-logy | Digital public services |
|---|---|---|---|---|---|---|---|
| 1. SII | Pearson's r | – | | | | | |
| | p-value | – | | | | | |
| 2. Connectivity | Pearson's r | 0.658 | – | | | | |
| | p-value | < 0.001 | – | | | | |
| 3. Human capital | Pearson's r | 0.530 | 0.647 | – | | | |
| | p-value | 0.003 | < 0.001 | – | | | |
| 4. Use of the internet | Pearson's r | 0.680 | 0.663 | 0.705 | – | | |
| | p-value | < 0.001 | < 0.001 | < 0.001 | – | | |
| 5. Integration of digital technology | Pearson's r | 0.700 | 0.698 | 0.730 | 0.816 | – | |
| | p-value | < 0.001 | < 0.001 | < 0.001 | < 0.001 | – | |
| 6. Digital public services | Pearson's r | 0.606 | 0.614 | 0.564 | 0.603 | 0.623 | – |
| | p-value | < 0.001 | < 0.001 | 0.001 | < 0.001 | < 0.001 | – |

The low tolerance values and VIF scores between 2 and 5 in the coefficient table indicate strong multicollinearity (Table 5); therefore, principal component analysis was used.

Table 5

### Coefficients

| Model | | Unstandardised | Standard error | Standardised | t | p | Collinearity statistics | |
|---|---|---|---|---|---|---|---|---|
| | | | | | | | tolerance | VIF |
| $H_0$ | (Intercept) | 57.534 | 2.266 | | 25.392 | < 0.001 | | |
| $H_1$ | (Intercept) | 12.662 | 10.712 | | 1.182 | 0.249 | | |
| | Connectivity | 0.332 | 0.264 | 0.257 | 1.259 | 0.221 | 0.429 | 2.331 |
| | Human capital | –0.145 | 0.221 | –0.137 | –0.654 | 0.519 | 0.404 | 2.476 |
| | Use of the internet | 0.211 | 0.209 | 0.248 | 1.009 | 0.323 | 0.296 | 3.376 |
| | Integration of digital technology | 0.295 | 0.257 | 0.301 | 1.148 | 0.263 | 0.260 | 3.844 |
| | Digital public services | 0.144 | 0.139 | 0.190 | 1.036 | 0.311 | 0.532 | 1.880 |

However, PCA resulted in a one-factor component, including each of the five dimensions of I-DESI, suggesting that those dimensions are strongly interrelated. (KMO= 0.881; Bartlett's Test of Sphericity: Approx. Chi-Square=85.289, df=10, Sig.= 0.000). The total percentage of the variance explained was 73.468. Table 6 presents the component matrix, in which only one-factor components can be seen. There are strong correlations between the five I-DESI dimensions and component 1. It was the integration of digital technology dimension that most highly correlated with the estimated component, which suggests that it is the most powerful dimension in the digital transformation of the economy and society.

Table 6

### Component matrix

| Connectivity | 0.845 |
|---|---|
| Human capital | 0.853 |
| Use of internet | 0.889 |
| Integration of digital technology | 0.908 |
| Digital public services | 0.786 |

*Note:* 1 component extracted. Extraction method: PCA.

All but the connectivity variable are normally distributed; therefore, connectivity was omitted from the model, and only four predictors remained (Table 7).

Table 7

Descriptive statistics

|  | SII | Connectivity | Human capital | Use of the internet | Integration of digital technology | Digital public sevices |
|---|---|---|---|---|---|---|
| Valid | 29 | 29 | 29 | 29 | 29 | 29 |
| Missing | 0 | 0 | 0 | 0 | 0 | 0 |
| Mean | 57.534 | 59.759 | 39.690 | 45.517 | 44.379 | 56.310 |
| Std. deviation | 12.202 | 9.425 | 11.604 | 14.339 | 12.448 | 16.123 |
| Shapiro-Wilk | 0.965 | 0.915 | 0.972 | 0.960 | 0.948 | 0.931 |
| P-value of Shapiro-Wilk | 0.427 | 0.022 | 0.616 | 0.332 | 0.166 | 0.059 |
| Minimum | 33.800 | 40.000 | 19.000 | 19.000 | 19.000 | 24.000 |
| Maximum | 79.400 | 72.000 | 62.000 | 68.000 | 62.000 | 80.000 |

Multiple linear regression with the stepwise method was used to find the model with the highest explanation power, resulting in a model with only one independent variable, which was the integration of digital technology. Scatter plots (residuals vs. covariates and residuals vs. predicted) show that the linear relationship and homoscedasticity conditions were met. Standardised residuals histogram shows normal distribution (Figure 5).

Figure 5

**Residuals vs. predicted and standardised residuals histogram (SII and IDT)**

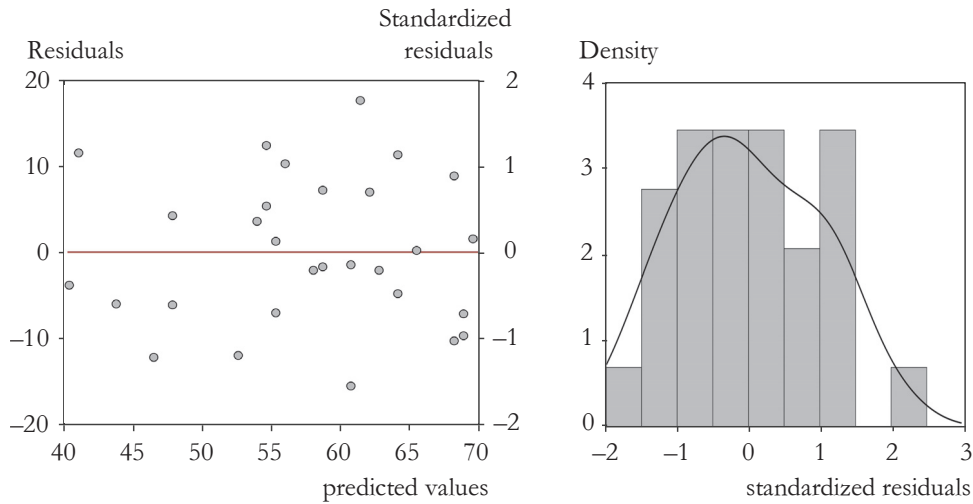

Linear regression analysis requires little or no autocorrelation in the data. The almost perfect Durbin-Watson's d value (1.988) signals no autocorrelation (Table 8).

Table 8
## Model summary (SII and IDT)

| Model | R | R² | Adjusted R² | RMSE | Durbin-Watson Auto-correlation | Durbin-Watson Statistic | Durbin-Watson p |
|---|---|---|---|---|---|---|---|
| $H_0$ | 0.000 | 0.000 | 0.000 | 12.202 | –0.165 | 2.214 | 0.559 |
| $H_1$ | 0.700 | 0.490 | 0.471 | 8.878 | –0.071 | 1.988 | 0.997 |

ANOVA results (Regression: sum of squares=2040.854, df=1, mean square=2040.854, F=25.894, p< .001) suggest that the model is significant and can be generalised.

The resulting model is robust; 49.0% of the variance in the SII is explained by the IDT variable. Pearson's correlation coefficient between the two variables (R=0.70) indicates a very strong relationship between social innovation capacity and IDT. The explanatory power of this model for the population is also high (47.1%). The social innovation capacity of a country can be defined as follows:

$$SII = 27.098 + 0.686 \times IDT$$

The standardised beta coefficient (β=0.700) suggests that IDT alone has a significant positive impact on social innovation capacity (Table 9).

Table 9
## Coefficients (SII and IDT)

| Model | | Unstan-dardised | Standard error | Standar-dised | t | p | Collinearity statistics tolerance | Collinearity statistics VIF |
|---|---|---|---|---|---|---|---|---|
| $H_0$ | (Intercept) | 57.534 | 2.266 | | 25.392 | < 0.001 | | |
| $H_1$ | (Intercept) | 27.098 | 6.204 | | 4.367 | < 0.001 | | |
| | Integration of digital technology | 0.686 | 0.135 | 0.700 | 5.089 | < 0.001 | 1.000 | 1.000 |

Pearson's correlation analysis between the five components of I-DESI and the SII revealed that all respective components are significantly, strongly, or moderately strongly associated with social innovation capacity. The strongest relationship was found between the SII and IDT, whereas the weakest relationship was observed between human capital and the SII (Table 10).

Table 10

**Pearson's correlations between SII and I-DESI dimensions**

| Variable | ° | SII | Connec-tivity | Human capital | Use of the internet | Integration of digital technology | Digital public sevices |
|---|---|---|---|---|---|---|---|
| 1. SII | Pearson's r | – | | | | | |
| 2. Connectivity | Pearson's r | 0.658*** | – | | | | |
| 3. Human capital | Pearson's r | 0.530** | 0.647*** | – | | | |
| 4. Use of the internet | Pearson's r | 0.680*** | 0.663*** | 0.705*** | – | | |
| 5. Integration of digital technology | Pearson's r | 0.700*** | 0.698*** | 0.730*** | 0.816*** | – | |
| 6. Digital public services | Pearson's r | 0.606*** | 0.614*** | 0.564** | 0.603*** | 0.623*** | – |

\* p < .05, \*\* p < .01, \*\*\* p < .001.

This finding suggests that the digitisation of business and the development of e-commerce have a serious impact on social innovations. In a country where the level of digitalisation of businesses is higher, the capacity for social innovation will also be higher. However, the skills required to take advantage of the opportunities offered by a digital society have the least impact on the social innovation potential.

The effect of use of the internet on social innovation capacity was also very strong. This suggests that in countries where people do many things online – making video calls, shopping, banking, and social networking – the capacity for social innovation will also be higher. Additionally, broadband infrastructure and its quality are strongly associated with the capacity for social innovation.

Very strong, positive correlations between each component of the I-DESI and civil society, the fourth pillar of SII were found (Table 11). This suggests that a more advanced digital society/economy makes a very serious positive impact on the culture of volunteerism and supports political participation. The increasing level of digitalisation of society enhances society engagement, the number of members (active or inactive) of a humanitarian or charitable organisation, and trust in society. The table also shows that many components of digitalisation positively influence the financing environment of social innovations.

Table 11

**Pearson's correlation table (I-DESI dimensions and the pillars of SII)**

| Variable | o | Connectivity | Human capital | Use of the internet | Integration of digital technology |
|---|---|---|---|---|---|
| 1. Connectivity | Pearson's r | – | | | |
| 2. Human capital | Pearson's r | 0.647*** | – | | |
| 3. Use of the internet | Pearson's r | 0.663*** | 0.705*** | – | |
| 4. Integration of digital technology | Pearson's r | 0.698*** | 0.730*** | 0.816*** | – |
| 5. Digital public services | Pearson's r | 0.614*** | 0.564** | 0.603*** | 0.623*** |
| 6. Policy and institutional framework | Pearson's r | 0.495** | 0.262 | 0.425* | 0.478** |
| 7. Financing | Pearson's r | 0.617*** | 0.494** | 0.683*** | 0.656*** |
| 8. Entrepreneurship | Pearson's r | 0.170 | 0.452* | 0.274 | 0.308 |
| 9. Society | Pearson's r | 0.666*** | 0.709*** | 0.788*** | 0.760*** |

| Variable | Digital public services | Policy and institutional framework | Financing | Entrepreneurship | Society |
|---|---|---|---|---|---|
| 1. Connectivity | | | | | |
| 2. Human capital | | | | | |
| 3. Use of the internet | | | | | |
| 4. Integration of digital technology | | | | | |
| 5. Digital public services | – | | | | |
| 6. Policy and institutional framework | 0.431* | – | | | |
| 7. Financing | 0.611*** | 0.631*** | – | | |
| 8. Entrepreneurship | 0.282 | 0.174 | 0.376* | – | |
| 9. Society | 0.579** | 0.355 | 0.738*** | 0.491** | – |

\* p < .05, \*\* p < .01, \*\*\* p < .001.

## Conclusions

Research findings confirmed that a country's higher digital performance leads to greater social innovation capacity. Consequently, more social innovation projects can be expected where the economy and society are digitally advanced. Therefore, decision-makers should consider accelerating the digital transformation to improve their country's capacity for social innovation (e.g., to increase the number of social innovation projects, create a better ecosystem for social innovations, etc.).

This research has significantly contributed to clarifying the relationship between digital society, the economy, and the capacity for social innovation. Based on the research findings, it can be concluded that digital competitiveness is a strong predictor of social innovation capacity and that social innovation potential can be estimated using the overall I-DESI score of a country.

It was also revealed that the integration of digital technology by businesses alone strongly influences social innovation capacity. This is the most crucial dimension of digital transformation. Therefore, the availability of the latest technologies and firm-level technology absorption should be supported to generate more social innovations. The number of SMEs selling online and the number of secure internet services should also be increased.

Digital transformation has a very strong positive impact on the development of civil society, the culture of volunteerism, and political participation. In a more digitally advanced country, we can expect higher levels of social engagement, an increase in the number of members of humanitarian or charitable organisations, and a rise in the level of trust in society.

As COVID-19 has accelerated almost all countries' digital transformation, it will have a favourable impact on their capacity for social innovation, from which society as a whole can benefit.

### Acknowledgement

This research was supported by the project nr. EFOP-3.6.2-16-2017-00007, titled „Aspects on the Development of Intelligent, Sustainable and Inclusive Society: Social, Technological, Innovation Networks in Employment and Digital Economy". The project has been supported by the European Union, co-financed by the European Social Fund and the budget of Hungary.

# APPENDIX

Table A1

### Raw dataset for analysis

|  | Country | SII | Policy and institutional framework | Financing | Entrepreneurship | Society | I-DESI | Connectivity | Human capital | Use of the internet | Integration of digital technology | Digital public services |
|---|---|---|---|---|---|---|---|---|---|---|---|---|
| 1 | USA | 79.4 | 84.6 | 80.4 | 76.2 | 68.4 | 59.0 | 70.0 | 55.0 | 49.0 | 50.0 | 68.0 |
| 2 | UK | 77.3 | 86.6 | 75.1 | 68.4 | 64.9 | 60.0 | 67.0 | 41.0 | 59.0 | 60.0 | 78.0 |
| 3 | Canada | 75.7 | 77.9 | 82.0 | 61.4 | 74.1 | 54.0 | 53.0 | 36.0 | 64.0 | 54.0 | 72.0 |
| 4 | Denmark | 71.2 | 67.2 | 75.6 | 61.1 | 83.9 | 65.0 | 72.0 | 55.0 | 68.0 | 62.0 | 69.0 |
| 5 | Belgium | 69.2 | 70.1 | 77.9 | 55.1 | 68.0 | 48.0 | 63.0 | 31.0 | 52.0 | 51.0 | 46.0 |
| 6 | New Zealand | 67.2 | 58.8 | 70.1 | 70.6 | 80.9 | 52.0 | 57.0 | 49.0 | 45.0 | 40.0 | 73.0 |
| 7 | France | 66.4 | 79.6 | 61.9 | 54.0 | 50.0 | 51.0 | 65.0 | 37.0 | 38.0 | 42.0 | 75.0 |
| 8 | Germany | 66.0 | 69.2 | 67.8 | 60.1 | 61.1 | 50.0 | 63.0 | 42.0 | 43.0 | 46.0 | 54.0 |
| 9 | Sweden | 65.7 | 63.3 | 69.3 | 60.8 | 71.2 | 59.0 | 70.0 | 45.0 | 61.0 | 56.0 | 65.0 |
| 10 | Switzerland | 61.6 | 55.8 | 69.0 | 58.4 | 69.2 | 57.0 | 70.0 | 52.0 | 57.0 | 61.0 | 41.0 |
| 11 | Australia | 60.6 | 49.2 | 72.9 | 59.6 | 74.2 | 60.0 | 64.0 | 50.0 | 60.0 | 52.0 | 80.0 |
| 12 | South Korea | 60.0 | 74.2 | 52.3 | 45.9 | 46.7 | 50.0 | 67.0 | 30.0 | 50.0 | 40.0 | 68.0 |
| 13 | Finland | 59.2 | 52.1 | 66.2 | 60.5 | 66.8 | 64.0 | 71.0 | 62.0 | 52.0 | 54.0 | 79.0 |
| 14 | Norway | 59.2 | 49.2 | 56.0 | 66.3 | 81.6 | 59.0 | 67.0 | 47.0 | 67.0 | 49.0 | 71.0 |
| 15 | Iceland | 59.0 | 52.1 | 51.4 | 55.0 | 88.3 | 61.0 | 67.0 | 60.0 | 64.0 | 61.0 | 49.0 |
| 16 | Netherlands | 57.7 | 47.2 | 52.9 | 75.9 | 74.3 | 59.0 | 66.0 | 49.0 | 55.0 | 60.0 | 68.0 |
| 17 | Italy | 57.5 | 64.2 | 54.1 | 51.4 | 50.4 | 39.0 | 59.0 | 25.0 | 24.0 | 39.0 | 43.0 |
| 18 | Chile | 56.9 | 65.4 | 43.8 | 67.7 | 43.1 | 45.0 | 47.0 | 48.0 | 37.0 | 46.0 | 41.0 |
| 19 | Ireland | 56.5 | 33.8 | 73.4 | 65.4 | 83.7 | 49.0 | 60.0 | 41.0 | 48.0 | 41.0 | 58.0 |
| 20 | Israel | 55.8 | 52.1 | 59.5 | 58.8 | 57.6 | 42.0 | 49.0 | 27.0 | 46.0 | 45.0 | 44.0 |
| 21 | Poland | 52.6 | 56.7 | 51.4 | 54.0 | 43.1 | 35.0 | 57.0 | 26.0 | 30.0 | 20.0 | 35.0 |
| 22 | Portugal | 52.0 | 53.8 | 38.2 | 61.1 | 56.6 | 38.0 | 59.0 | 31.0 | 30.0 | 30.0 | 36.0 |
| 23 | Japan | 48.0 | 49.1 | 54.4 | 44.8 | 40.4 | 49.0 | 66.0 | 37.0 | 44.0 | 41.0 | 54.0 |
| 24 | Spain | 44.8 | 41.7 | 44.8 | 52.3 | 46.2 | 50.0 | 60.0 | 37.0 | 36.0 | 49.0 | 72.0 |
| 25 | Russia | 41.4 | 46.3 | 38.2 | 46.3 | 29.2 | 38.0 | 42.0 | 36.0 | 46.0 | 30.0 | 38.0 |
| 26 | Mexico | 40.2 | 36.7 | 39.8 | 50.5 | 40.9 | 34.0 | 43.0 | 24.0 | 19.0 | 37.0 | 49.0 |
| 27 | Brazil | 37.4 | 28.8 | 41.3 | 59.4 | 35.4 | 37.0 | 48.0 | 35.0 | 26.0 | 24.0 | 46.0 |
| 28 | Turkey | 36.2 | 30.9 | 36.9 | 64.9 | 24.5 | 26.0 | 40.0 | 19.0 | 21.0 | 19.0 | 24.0 |
| 29 | China | 33.8 | 29.2 | 37.5 | 53.9 | 24.0 | 34.0 | 51.0 | 24.0 | 29.0 | 28.0 | 37.0 |

*Source:* EIU (2016, p. 11).